\documentclass{myarticle}
\usepackage{times}
\usepackage{graphicx}
\usepackage{latexsym}
\usepackage{epic}
\usepackage{amsfonts}
\usepackage{amsmath}
\usepackage{amssymb}
\usepackage{amsthm}
\usepackage{xspace}
\usepackage{mathrsfs}
\usepackage{subfigure}
\usepackage{newalg}

\begin{document}

\newtheorem{theorem}{Theorem}
\newtheorem{lemma}{Lemma}
\newtheorem{definition}{Definition}
\newtheorem{myexample}{Example}
\newtheorem{mytheorem}{Theorem}
\newcommand{\myproof}{\noindent {\bf Proof:\ \ }}
\newcommand{\myqed}{\mbox{$\Box$}}
\newcommand{\nmax}{N}
\newcommand{\zuck}{\mbox{\sc Reverse}}
\newcommand{\lslg}{\mbox{\sc Largest Fit}}
\newcommand{\lsla}{\mbox{\sc Average Fit}}
\newcommand{\lslgbf}{\mbox{\bf Largest Fit}}
\newcommand{\lslabf}{\mbox{\bf Average Fit}}
\newcommand{\myOmit}[1]{}

\newcommand{\mf}{{\mathscr F}}
\newcommand{\mc}{\mathcal C}
\newcommand{\ra}{\rightarrow}

\newcommand{\ms}{\mathcal S}
\newcommand{\ma}{\mathcal A}
\newcommand{\mv}{\mathcal V}
\newcommand{\rev}{\text{rev}}
\newcommand{\others}{\text{\it Others}}

\newcommand{\vote}[3]{\mbox{$#1 \! \succ \! #2 \! \succ \! #3$}\xspace}
\newcommand{\vvote}[4]{\mbox{$#1 \! \succ \! #2 \! \succ \! #3 \! \succ \! #4$}\xspace}

\newcommand{\reverse}{\mbox{\sc Reverse}\xspace}
\newcommand{\largestfit}{\mbox{\sc LargestFit}\xspace}
\newcommand{\averagefit}{\mbox{\sc AverageFit}\xspace}
\newcommand{\eliminate}{\mbox{\sc Eliminate}\xspace}
\newcommand{\reveliminate}{\mbox{\sc RevEliminate}\xspace}

\newcommand{\eliminatef}{\mbox{\bf Eliminate}}
\newcommand{\reveliminatef}{\mbox{\bf Reverse Eliminate}}

\title{Generating Single Peaked Votes}


\author{Toby Walsh}
\address{NICTA and UNSW, Sydney, Australia, toby.walsh@nicta.com.au}

\begin{abstract}
We discuss how to generate singled peaked
votes uniformly from the Impartial Culture model.
\end{abstract}

\maketitle
\sloppy

\section{Introduction}

Within computational social choice,
there is increasing attention away from the worst-case
complexity of manipulation and control of voting rules,
and focus on performance on average (for example, 
\cite{praamas2007,xcec08,fknfocs09,wijcai09,wecai10,wjair11})
and in practice (using, for example, resources
like PrefLib \cite{preflib}). There is also
an increasing focus on identifying tractable
special cases using tools like
fixed parameter tractability (e.g. \cite{butcs09,hlmcorr2014})
and domain restrictions 
(e.g. \cite{waaai2007,fhhrtark09}).
One of the most common domain restrictions
considered in social choice theory is that
of single peaked preferences. 
In this note, we discuss how to sample 
single peaked votes uniformly, correcting
a common mistake found in the literature.

\section{Generating single peaked votes}

We suppose that the candidates can be ordered
from left to right. An
agent's preferences are single peaked
when the agent prefers alternatives closer to their
most preferred candidate. 
How do we generate single peaked votes
uniformly, that is, under the Impartial Culture (IC) model?
Suppose candidates are represented by
integer points from 1 to $n$ and
votes are single peaked along this line. 
For example, for $n=3$,
we have 4 different single peaked votes:
$1>2>3$, 
$2>1>3$, 
$2>3>1$ and
$3>2>1$. 
Notice that half of these single peaked 
votes end in 3 and, excluding the last ranked candidate 3,  
are themselves single peaked votes from 1 to 2. 
On the other hand, the other half of these
single peaked votes end in 1 and, excluding the 
last ranked candidate 1, are single peaked votes from 2 to 3. 

In general, given candidates at integer points on the line 
1 to $n$, there are $2^{n-1}$ different singled peaked votes.
Half of all these single peaked votes end in
$n$ and are made up of all the single peaked votes from 1 to $n-1$ 
augmented with $n$ at their end. The other half of these single peaked votes
end in $1$ and are made up of all the single peaked votes from 2 to $n$
augmented with $1$ at their end. This motivates the following 
recursive procedure which 
uniformly generates one of the $2^{b-a}$
possible single peaked votes
in the interval $[a,b]$. The vote
is returned as a rank ordered list of integers. 

~ \\

\begin{algorithm}{GenSinglePeak}{a,b}
\begin{IF}{a=b}
   {\RETURN []}
\ELSE
\begin{IF}{coin-toss=heads}
  {\RETURN
    Append(\CALL{GenSinglePeak}(a+1,b),[a])}
\ELSE
  {\RETURN
    Append(\CALL{GenSinglePeak}(a,b-1),[b])}
\end{IF}
\end{IF}
\end{algorithm}

~ \\

Other methods have been proposed to generate
singled peaked votes that sample from a different
distribution. 
For example, Conitzer generated single peaked
preferences by randomly picking a peak,
and then randomly choosing the next highest alternative
to the left or right of the positions currently ranked \cite{cjair09}. 
With 3 candidates, 
there is a $\frac{1}{3}$ chance that Conitzer's
generator will return the votes
$2>1>3$ or $2>3>1$, but the IC
model has a $\frac{1}{2}$ chance to generate these votes. 
Similarly, there is 
a $\frac{1}{3}$ chance that Conitzer's
generator will return the vote
$1>2>3$ (and a $\frac{1}{3}$ chance that 
it will return the vote $3>2>1$),
but the IC model has only a $\frac{1}{4}$ chance to generate this vote. 
More generally, the IC model is more likely
to return votes with peaks in the middle of
the left-right spectrum than Conitzer's
generator. On the other hand, Conitzer's
generator is more likely to returns votes with peaks at
the ends of the left-right spectrum than the IC model. 
Asymptotically, the difference between the two
models is extreme. For example, the probability that
the IC model returns the single peaked vote
$1 > \ldots > n$ is exponentially small ($\frac{1}{2^n}$)
whilst Conitzer's generator returns
this vote with a much greater probability ($\frac{1}{n}$). 
Note that as $n$ goes to infinity, the ratio
between these two probabilities goes to zero.

\section{Acknowledgements}

Toby Walsh is supported by the
Australian Department of Communications, 
the ARC, and the Asian Office of Aerospace Research and
Development. 



%
\bibliographystyle{plain}
\bibliography{/Users/twalsh/Documents/biblio/a-z,/Users/twalsh/Documents/biblio/a-z2,/Users/twalsh/Documents/biblio/pub,/Users/twalsh/Documents/biblio/pub2}
\end{document}